\magnification=1200
\def\bk{{\bf k}}
\def\bp{{\bf p}}
\def\ng{N_g}
\def\ngu{N_{g_1}}
\def\ngd{N_{g_2}}
\def\ngt{N_{g_3}}
\def\bu{{\bf u}_g}
\def\ep{\epsilon}
\def\bb{{\bf b}}
\def\bx{{\bf x}}
\def\bv{{\bf v}}
\def\be{{\bf E}}

\def\np{n_{{\bf p}}}
\def\nn{{\rm N}}

\def\cps{\Psi}

\def\cph{\Phi}
\def\ps{\psi}
\def\ph{\varphi}
\def\dd{{\cal D}}
\def\hh{{\cal H}}
\def\al{\alpha}

\def\phg{\Phi_g}
\def\phgu{\Phi_{g_1}}
\def\phgd{\Phi_{g_2}}
\def\phgt{\Phi_{g_3}}
\def\psg{\Psi_g}
\def\psgu{\Psi_{g_1}}
\def\psgd{\Psi_{g_2}}
\def\psgt{\Psi_{g_3}}
\def\ogu{\omega_{g_1}}
\def\ogd{\omega_{g_2}}
\def\ogt{\omega_{g_3}}
\def\oo{\omega_g}

\centerline{\bf GENERALIZED KINETIC THEORY OF ELECTRONS AND PHONONS }
 
\vskip .3cm
\centerline{A. Rossani}
\centerline{Istituto Nazionale di Fisica della Materia,}
\centerline{Dipartimento di Fisica,
Politecnico di Torino,}
\centerline{Corso Duca degli Abruzzi 24, 43100 Torino, Italy}
\vskip .3cm
{\bf Abstract.} A Generalized Kinetic Theory [1] was proposed in order
to have the possibility to treat particles which obey a very general
statistics. By adopting the same approach, we generalize here the
Kinetic Theory of electrons and phonons.
Equilibrium solutions and their stability are investigated. 
\vskip .3cm

{\bf 1. Introduction}

Very recently [1], a Generalized Kinetic Theory (GKT) has been proposed
by Rossani and Kaniadakis, in order to have the possibility to treat, at
a
kinetic level, particles which obey a very general statistics.

The quasi--classical Boltzmann equation introduced in ref.
[1]
is a generalization of the Uehling--Uhlenbeck equation [2], which is
intended
for bosons and fermions only. 
In ref. [1] we have shown that our generalized Uehling--Uhlenbeck
equation (GUUE) assures particle, momentum and energy conservation.
Equilibrium, its uniqueness and stability
(via an H--theorem) have been investigated.

In ref. [3] Koponen points out that fractal or inverse power law
distributions of phonon excitations are of interest in modeling
various meaningful situations in solid state physics. 
Moreover, he feels that until recently there has been
little guidance on how to generalize
the kinetic theory of electrons and phonons obeying non Gibbsian
statistics.

Here we propose a Generalized Kinetic Theory for Electrons and 
Phonons (GKTEP), by following the same ideas which lead to the GUUE.
In order to keep our GKTEP as general as possible,   
we introduce modified collision terms not only for phonons, but also
for electrons, 
so that an application is allowed not only to electrons, but also
to other particles (obeying a general statistics) which interact with a crystal lattice.

The paper is organized as follows. 
In section 2 the Bloch--Boltzmann--Peierls [4] equations are briefly
recalled, and our
generalization is introduced.

In section 3 and 4
 we study the equilibrium solutions to the GKTEP equations and their
stability is investigated, via an H theorem.

\vskip .3cm
{\bf 2. Generalized kinetic equations for electrons and phonons}

The most complete description, at a {\it mesoscopic} level, of a system
of electrons and phonons, is based on the Bloch--Boltzmann--Peierls
(BBP) 
equations [4].

Consider two populations: electrons (e), whose number is conserved, obey Fermi--Dirac
statistics; phonons (p), whose number is not conserved, obey 
Bose--Einstein statistics. 
Of course, one should deal, in principle, with p--p, p--e, and e--e
interactions.
A first important assumption is the following: electrons (distribution
function $n$) are considered
as a rarefied gas in a "sea" of phonons (distribution function
$N$).
This means that e--e interactions can be neglected.
In e--p interactions only electrons and energy are conserved. In p--p
interactions only energy is conserved.

By adopting the notation of ref. [5], 
let $\ng=\ng(\bk,\bx,t)$ be the distribution function of phonons
(quasi--momentum $\bk$, energy $\oo(\bk)$) of type
$g$
(i.e. branch $g$ of the phonon spectrum) and $n=n(\bp,\bx,t)$ the
distribution function of electrons (quasi--momentum $\bp$, energy $\ep(\bp)$).
The distribution functions $\ng$ and $n$ are normalized so that
the thermal energy density $E_p$ of the crystal is given by
$$E_p(\bx,t)={1\over 8\pi^3}\sum_g\int\oo \ng(\bk,\bx,t)d\bk,
$$
while the concentration $\nn$ and the energy density
$E_e$ of the electron gas are given, respectively by
$$\nn(\bx,t)={1\over 8\pi^3}\int n(\bp,\bx,t)2 d\bp,\ \ E_e(\bx,t)={1\over
8\pi^3}\int\ep(\bp)n(\bp,\bx,t)2d\bp,$$
where the factor 2 inside these integrals accounts for degeneracy.

The BBP equation for phonons reads
$${\partial \ng\over\partial t}+\bu\cdot{\partial\ng\over\partial\bx}
=\ \left({\partial\ng\over\partial t}\right)_{coll}
=\left({\partial \ng\over\partial t}\right)_{pp}+\left({\partial
\ng\over\partial t}\right)_{pe}\eqno(1)$$
($\bu={\partial\oo/\partial\bk}$), where
$$\eqalign{&\left({\partial\ng\over\partial t}\right)_{pp}=\int\lbrace
-(1/2)\sum_{g_1g_2}w_{pp}(\bk_1,\bk_2\rightarrow\bk)\delta(\oo-\ogu-\ogd)
[\ng(\ngu+1)(\ngd+1)\cr &-(\ng+1)\ngu\ngd]
+\sum_{g_1g_3}w_{pp}(\bk,\bk_1\rightarrow\bk_3)\delta(\ogt-\oo-\ogu)
[(\ng+1)(\ngu+1)\ngt\cr &-\ng \ngu(\ngt+1)]\rbrace{d\bk_1\over 8\pi^3}\cr}$$
and 
$$\eqalign{&\left({\partial \ng\over\partial t}\right)_{pe}=\cr &\int
2w_{pe}(\bp\rightarrow\bp',\bk)
[n_{\bp}(1-n_{\bp'})(1+\ng)-n_{\bp'}(1-n_{\bp})\ng]\delta(\ep_{\bp'}+\oo-\ep_{\bp}){d\bp\over
8\pi^3},\cr}$$
where $\np=n(\bp)$, $\ep_{\bp}=\ep(\bp)$, $w_{pp}$ 
and $w_{pe}$ are transition probabilities.

The first term in the braces corresponds to the direct and reverse processes
$$(g,\bk)\ \rightleftharpoons\ (g_1,\bk_1)\ +\ (g_2,\bk_2),$$
where $$\bk_2=\bk-\bk_1-\bb\eqno(2)$$ 
and $\bb$ is a vector of the reciprocal lattice.

The second term in braces correspond to the processes
(direct and reverse) 
$$(g_3,\bk_3)\ \rightleftharpoons\ (g,\bk)\ +\ (g_1,\bk_1),$$
where $$\bk_3=\bk+\bk_1+\bb.\eqno(3)$$

The BBP equation for electrons reads
$${\partial\np\over\partial t}+\bv\cdot{\partial
\np\over\partial\bx}
-e\be\cdot{\partial\np\over\partial\bp}=
\left({\partial n_{\bp}\over\partial t}\right)_{ep}\eqno(4)$$
(e--e interactions have been neglected), where  $-e$ is the electron
charge and
$$\eqalign{&\left({\partial n_{\bp}\over\partial t}\right)_{ep}=\sum_g\int\lbrace w_{ep}(\bp',\bk\rightarrow\bp)[n_{\bp'}(1-n_{\bp})\ng
\cr &-n_{\bp}(1-n_{\bp'})(1+\ng)]\delta(\ep_{\bp}+\ep_{\bp'}-\oo)\cr
&+w_{ep}(\bp'\rightarrow\bp,\bk)[n_{\bp'}(1-n_{\bp})(1+\ng)
-n_{\bp}(1-n_{\bp'})\ng]\delta(\ep_{\bp}+\oo-\ep_{\bp'})\rbrace{d\bk\over
 8\pi^3},\cr}$$
where $\np=n(\bp)$, $\bv={\partial\ep_{\bp}/\partial\bp}$,
and the transition probabilities $w_{ep}$'s obey the following relationships:
$$w_{pe}(\bp\rightarrow\bp',\bk)=w_{ep}(\bp\rightarrow\bp',\bk)=w_{ep}(\bp',\bk\rightarrow\bp).$$

The first term corresponds to processes with emission of a phonon having
quasi--momentum $\bk$
by an electron having a given quasi--momentum $\bp$, and reverse
processes with absorption of a phonon $\bk$ by electrons $\bp'$ with
return to the quasi--momentum $\bp$:
$$\bp=\bp'+\bk+\bb.\eqno(5)$$

The second term corresponds to processes with absorption of a phonon
by an electron $\bp$ and the reverse processes of its emission by
electrons $\bp'$:
$$\bp+\bk=\bp'+\bb.\eqno(6)$$

We shall generalize [1] the expressions of 
$$\left({\partial \ng\over\partial t}\right)_{pp},\ \ \left({\partial \ng\over\partial t}\right)_{pe}, 
\ \ \left({\partial \np\over\partial t}\right)_{ep},$$ 
by introducing the following substitutions:
$$\eqalign{&1+\ng\ \rightarrow\ \cps(\ng),\ \ \ng\ \rightarrow\ \cph(\ng)\cr
&1-\np\ \rightarrow\ \ps(\np),\ \ \np\ \rightarrow\ \ph(\np),\cr}$$
where
$\cps$, $\cph$, and $\ps$, $\ph$, are, respectively, non negative
functions of $\ng$ and $\np$, obeying the conditions
$$\cps(0)=\ps(0)=1,\ \ \ \cph(0)=\ph(0)=0.$$
We assume that $\cph(\ng)/\cps(\ng)$ and $\ph(\np)/\ps(\np)$ are
monotonically increasing functions, respectively, of $\ng$ and $\np$.
This is trivially true for bosons and fermions. In general, this
assumption is justified {\it a posteriori}, since it assures uniqueness
and stability of equilibrium. 

The functionals $\ps,\ \ph$ and $\cps,\ \cph$ replace the Pauli bloking
and boson enhancement factors in the phase space, respectively. 
Alternatively, one could say that generalized statistics, defined by
means of the couples $\cps,\ \cph$ and $\ps,\ \ph$, are introduced 
for phonons and electrons, respectively.

We have then
$$\eqalign{&\left({\partial\ng\over\partial t}\right)_{pp}=-(1/2)\int\lbrace
\sum_{g_1g_2}w_{pp}(\bk_1,\bk_2\rightarrow\bk)\delta(\oo-\ogu-\ogd)
[\cph(\ng)\cps(\ngu)\cps(\ngd)-\cr &\cps(\ng)\cph(\ngu)\cph(\ngd)]+
\sum_{g_1g_3}w_{pp}(\bk,\bk_1\rightarrow\bk_3)\delta(\ogt-\oo-\ogu)
[\cps(\ng)\cps(\ngu)\cph(\ngt)-\cr &\cph(\ng)\cph(\ngu)\cps(\ngt)]\rbrace{d\bk_1\over 8\pi^3},\cr}$$
$$\eqalign{&\left({\partial \ng\over\partial t}\right)_{pe}=\cr &\int
2w_{pe}(\bp,\bp'\rightarrow\bk)
[\ph(n_{\bp})\ps(n_{\bp'})\cps(\ng)-\ph(n_{\bp'})\ps(n_{\bp})\cph(\ng)]\delta(\ep_{\bp'}+\oo-\ep_{\bp}){d\bp\over
8\pi^3},\cr}$$
$$\eqalign{&\left({\partial n_{\bp}\over\partial t}\right)_{ep}
=\sum_g\int\lbrace w_{ep}(\bp',\bk\rightarrow\bp)[\ph(n_{\bp'})\ps(n_{\bp})\cph(\ng)
\cr &-\ph(n_{\bp})\ps(n_{\bp'})\cps(\ng)]\delta(\ep_{\bp}+\ep_{\bp'}-\oo)\cr
&+w_{ep}(\bp'\rightarrow\bp,\bk)[\ph(n_{\bp'})\ps(n_{\bp})\cps(\ng)
-\ph(n_{\bp})\ps(n_{\bp'})\cph(\ng)]\delta(\ep_{\bp}+\oo-\ep_{\bp'})\rbrace{d\bk\over
 8\pi^3}.\cr}$$
\vskip .3cm
{\bf 3. Equilibrium}

Even though the generalized equations are certainly more complicated
that the original ones, many of the usual techniques of kinetic theory
are still applicable in the study of equilibria and their stability.
In particular, we will prove an H theorem for the present problem, whose
connections with the entropy law are easily established.

In the space homogeneous and forceless case, equilibrium is defined by 
$$\left({\partial\ng\over\partial
t}\right)_{coll}=\left({\partial\np\over\partial t}\right)_{ep}=0.\eqno(7)$$ 
In order to study equilibrium and its stability, it is useful to introduce the following functional:
$$\dd=\sum_g\int\left({\partial\ng\over\partial
t}\right)_{coll}\ln{\cph(\ng)\over\cps(\ng)}
d\bk+2\int\left({\partial\np\over\partial
t}\right)_{ep}\ln{\ph(\np)\over\ps(\np)}
d\bp,\eqno(8)$$
which can be put in the following form:
$$\eqalign{&\dd={1\over2}\ \sum_{g_1g_2g_3}\int\int 
 w_{pp}(\bk_2,\bk_3\rightarrow\bk_1)
\delta(\ogu-\ogd-\ogt)
(-\phgu\psgd\psgt+\cr &+\psgu\phgd\phgt)\ln{\phgu\psgd\psgt\over\psgu\phgd\phgt}d\bk_1d\bk_2
 +2\sum_g\int\int w_{ep}(\bp\rightarrow\bp',\bk)\times\cr
&\times(-\ph'\ps\phg+\ph\ps'\psg)\delta(\ep_{\bp'}+\oo-\ep_{\bp})\ln{\ph'\ps\phg\over\ph\ps'\psg}d\bk 
d\bp\ \ \le\ 0,\cr}\eqno(9)$$
where
$$\eqalign{&\Phi_{g_i}=\Phi(\ng(\bk_i)),
\ \ \Psi_{g_i}=\Psi(\ng(\bk_i)),\cr
&\varphi'=\varphi(n_{\bp'}),\ \ \psi'=\psi(n_{\bf p'}).\cr}$$

{\bf Proposition 1.}

{\it Condition $(7)$ is equivalent to the following couple of equations:}
$$\eqalignno{&\cps(\ng)\cph(\ngu)\cph(\ngd)=\cph(\ng)\cph(\ngu)\cps(\ngd)\
 \ \ 
\forall\ \bk,\ \bk_1 &(10)\cr
&\ph(n_{\bp})\ps(n_{\bp'})\cps(\ng)=\ph(n_{\bp'})\ps(n_{\bp})\cph(\ng)
\ \ \ \forall\ \bp,\ \bk 
.&(11)\cr}$$

{\bf Proof.} First of all we observe that $(10),\ (11)\ \Longrightarrow\
(7)$.
On the other hand, from (8) we have $(7)\ \Longrightarrow\ \dd=0$. Now, since
in eq. (9) both the integrands are never positive, one has $\dd=0\
\Longrightarrow\ (10),\ (11)\ \bullet$

Condition (10) shows that $\ln(\cph^*/\cps^*)$ is a
collisional invariant for phonons, that is
$$\ln{\cph(\ng^*)\over\cps(\ng^*)}=-\oo/T\eqno(12)$$
(hereinafter $\ast$ means "at equilibrium"), where $T$ is the absolute
temperature of the whole system (electrons plus phonons).

From (11) and (12), taking into account that $\ep_{\bp}=\ep_{\bp'}+\oo$, we find
that, at equilibrium, 
$\ln[\ph(\np)/\ps(\np)]+\ep_{\bp}/T$ is a collisional invariant
for electrons, that is
$$\ln{\ph(\np^*)\over\ps(\np^*)}=(\mu-\ep_{\bp})/T,\eqno(13)$$
where $\mu$ is the chemical potential of the electron gas.

Observe that, due to the monotonicity of both $\cph/\cps$ and $\ph/\ps$,
equations (12) and (13) give unique solutions for $\ng^*$ and $\np^*$,
respectively.

We would like to stress that, with a proper choice of the ratios
$\cph/\cps$
and $\ph/\ps$, it is possible to reproduce the non-standard quantum
statistics proposed in the last years and compare them from the point of view
of the GKTEP, but this will be matter of a separate future paper [6].

\vfill\eject 
{\bf 4. Stability}

In order to study the stability of such equilibrium solutions,
let us introduce the following functional:
$$H=H_p+H_e=\sum_g\int\hh_p(\ng)d\bk+\int\hh_e(\np)d\bp,$$
where
$${\partial\hh_p(\ng)\over\partial\ng}=\ln{\cph(\ng)\over\cps(\ng)},\ \ 
{\partial\hh_e(\np)\over\partial\np}=2\ln{\ph(\np)\over\ps(\np)}.$$
Observe that, since $\cph/\cps$ and $\ph/\ps$ have been assumed to be
monotonically increasing, $\hh_p$ and $\hh_e$ are convex functions of 
$\ng$ and $\np$, respectively.

We can now prove the following H theorem:

{\bf Proposition 2.}

{\it $H$ is a Lyapounov functional for the present problem.}

{\bf Proof.} First of all, it is easily verified that $\dot H=\dd\le0$.
Then we observe that
$$\sum_g\int\left({\partial\hh_{pg}\over\partial\ng}\right)^*(\ng-\ng^*)d\bk
+\int\left({\partial\hh_{e\bp}\over\partial\np}\right)^*(\np-\np^*)d\bp=0,$$
where $\hh_{pg}=\hh_p(\ng)$ and $\hh_{e\bp}=\hh_e(\np)$,
due to electron and energy conservation.

Now we can write
$$H-H^*=\sum_g\int\hat\hh_{pg}d\bk+\int\hat\hh_{e\bp}d\bp,$$
where
$$\hat\hh_{\al}=\hh_{\al}-\left[\hh_{\al}^*
+\left({\partial\hh_{\al}\over\partial\phi_{\al}}\right)^*
(\phi_{\al}-\phi_{\al}^*)\right],$$
where $\phi_{\al}=\ng$ for $\al=pg$ and $\phi_{\al}=\np$ for $\al=e\bp$.

Due to the convexity of $\hh_{\al}$, we have 
$\hat\hh_{\al}\ \ge\ 0$ and finally $H-H^*\ \ge\ 0\ \bullet$  

By definining entropy as $$S=S_p+S_e=-{H_p+H_e\over 8\pi^3},$$ these results are easily
interpreted on a physical ground. In fact, $\dot H\le0$ and
 $H-H^*\ge0$ simply mean that entropy is always non decreasing and
it attains a maximum at equilibrium.

Observe that this definition of $S$ is consistent with
the definitions of $T$ and $\mu$ we already gave. In fact, at equilibrium, the following 
thermodynamical relationships are recovered:
$$\left({\partial S\over\partial\nn}\right)_E=-{\mu\over T},\ \ \left({\partial
S\over\partial E}\right)_\nn={1\over T},$$
where $E=E_p+E_e$.
\vskip .3cm
{\bf AKNOWLEDGEMENT}

This work has ben supported by the Fond zur Foerderung der
wissenshaftlinken
Forscung, Vienna, under contract No. P14669-TPH
\vfill\eject

{\bf References}

[1] A. Rossani, G. Kaniadakis, "A generalized quasi--classical Boltzmann
equation",
{\it Physica A}, 277 (2000) 349 

[2] R. L. Liboff, {\bf Kinetic Theory}, Prentice Hall, London (1990)

[3] I. Koponen, "Thermalization of an electron--phonon system in a
nonequilibrium
state characterized by fractal distribution of phonon excitations",
{\it Phys. Rev. E}, Vol. 55, No. 6, 7759 (1997)

[4] J. M. Ziman, {\bf Electrons and Phonons}, Claredon Press, Oxford (1967)

[5] E. M. Lifshitz, L. P. Pitaevskii, {\bf Physical Kinetics}, Pergamon
Press,
Oxford (1981)

[6] A. Rossani, A.M. Scarfone, submitted to {\it Physica A}, 2001

\end